\documentstyle[12pt,a4]{article}
\title{Production of three Vector Bosons\\in $\rm e^+e^-$ Annihilation\\
as a Test of $\rm W^\pm$, Z, $\gamma$ Self-Interactions}
\author{C. Grosse-Knetter\thanks{Partially supported by Deutsche
Forschungsgemeinschaft}\\
and\\ D. Schildknecht\\[0.5cm]
Universit\"at Bielefeld\\ Fakut\"at f\"ur Physik\\
D-4800 Bielefeld 1\\ Germany}
\date{BI-TP 92/30\\July 1992\\(Enlarged Version October 1992)}

\newcommand{\dk}{\Delta\kappa}
\newcommand{\eref}[1]{(\ref{#1})}
\newcommand{\lag}{{\cal L}}
\newcommand{\wwz}{$\rm e^+e^-\to W^+W^-Z$}
\newcommand{\wwzlll}{$\rm e^+e^-\to W^+_LW^-_LZ^{ }_L$}
\newcommand{\wwa}{$\rm e^+e^-\to W^+W^-\gamma$}
\newcommand{\ww}{$\rm e^+e^-\to W^+W^-$}
\newcommand{\ep}{$\rm e^+e^-$}
\newcommand{\be}{\begin{equation}}
\newcommand{\ee}{\end{equation}}
\newcommand{\bea}{\begin{eqnarray}}
\newcommand{\eea}{\end{eqnarray}}
\newcommand{\suu}{$\rm SU(2)_L\times U(1)_Y$}
\newcommand{\suwi}{$\rm SU(2)_{WI}$}
\newcommand{\ws}{\protect\sqrt{s}}
\newcommand{\gev}{\,\rm GeV}
\newcommand{\fb}{\,\rm fb}
\newcommand{\wza}{$\rm W^\pm$, Z, $\gamma$}
\newcommand{\tw}{\theta_W}
\newcommand{\kp}{\kappa}
\newcommand{\nn}{\nonumber\\}
\newcommand{\st}{\sigma_{tot}}
\newcommand{\wmu}{$\vec{W}_\mu$}
\newcommand{\bfw}{$\vec{W}$}
\newcommand{\owp}{{\cal O}_{W\Phi}}
\newcommand{\sw}{\sin\tw}

\newcommand{\cwq}{\cos^2\tw}

\begin{document}
\setcounter{page}{0}
\thispagestyle{empty}
\maketitle
\begin{abstract}
We study the vector-boson production
processes \wwz\ and \wwa\  which are directly affected
by the trilinear and quadrilinear
self couplings of the $\rm W^\pm$, Z and $\gamma$.
Our analysis is based upon a single-parameter effective-Lagrangian
model for these self interactions which
contains the standard model as a special case.
Consequences for the phenomenology at an \ep  collider
of 500$\gev$ (NLC) are discussed, and fits of the free parameter around its
standard model value are carried out.
\end{abstract}
\newpage


\section[ ]{Introduction}
The standard \suu\  model of electroweak interactions \cite{sm}
has been confirmed by all
experiments up to now. However, this empirical evidence is essentially
restricted to vector-boson--fermion interactions. The vector-boson
self interactions, which are a
consequence of the non-Abelian structure of the gauge group \suu\
(in its minimal realization), and which are
essential for the renormalizability and unitarity  of the theory, contribute
to reactions realized at present collider energies only indirectly via
vector-boson-loop corrections.

Future colliders like LEP II (\ep\ at $\ws=200\gev $) and NLC
(\ep at $\ws=500\gev $) \cite{nlc}
will make direct tests of \wza\ self interactions
possible by the measurement of processes which get tree level
contributions from
these self couplings. The mainly studied process of this type is the
two-vector-boson production
process \ww\ (e.g.\ \cite{gg,schi4,schi1}) as a test of the
trilinear self couplings. A full confirmation of the non-Abelian vector-boson
sector of the model, however, needs a test of the quadrilinear couplings
as well.
One class of processes which is directly effected by tri- and
quadrilinear self couplings, is vector-boson scattering
($\rm V_1,V_2\to V_3,V_4$
with $V_i=W^\pm,Z,\gamma$) \cite{schi2,schi3}, but, unfortunately, it
can only be
measured in reactions of the type $\rm e^+e^-\to(e^+e^-,\bar{\nu}e^-,\nu e^+,
\nu\bar{\nu})+V_1V_2$ at  CM-energies in the TeV region,
which cannot be realized experimentally yet.

The reactions with direct contributions of tri- and quadrilinear self couplings
which are of greatest phenomenological interest are the
three-gauge-boson production
processes \wwz\ and \wwa\ \cite{bh,tnk,wwz},
which will be measurable
at the expected new linear collider (NLC) with a CM-energy of $\ws=500\gev $
and a luminosity of $\rm 20\, fb^{-1}a^{-1}$. The standard model cross sections
for these processes were calculated by  Barger, Han and Phillips \cite{bh}
and by Tofighi-Niafi and Gunion \cite{tnk}. To examine the sensitivity of
these cross sections to deviations of the self couplings from
their standard model
values and to find experimental bounds on such deviations,
we need a more general parameter-dependent model which includes the
standard model as a special case. The most general procedure would
start from an effective Lagrangian containing all
vector-boson self interactions which can be constructed
in agreement with Lorentz invariance. Such an analysis was
performed for trilinear self couplings and the $\rm W^+W^-$-production
process in \cite{gg} and \cite{schi4}. For three-vector-boson production,
however, this seems to be a complicated procedure
involving elaborate multi-parameter fits and large individual errors
of the fit parameters as a consequence of limited future statistics.

For simplification we base our analysis on the less extended KMSS model
\cite{schi1,schi2}, which rests upon global \suwi\
symmetry broken by
electromagnetism. Electromagnetic interactions are assumed to be $P$- and $C$-
invariant and only dimensionless coupling constants are allowed,
thus suppressing
dimension-six quadrupole terms. These assumptions imply a four-parameter
Lagrangian; the free parameters can be reduced to a single one, which is
chosen as the anomalous magnetic moment $\kappa$ of the $\rm W^\pm$ boson
if, in addition, one imposes the requirement that those terms in the
vector-boson scattering amplitudes which grow most strongly with energy
(as $s^2$) are absent (BKS model) \cite{schi2}.

It turns out, that this single-parameter model (extended
by appropriate couplings of the Higgs boson to the vector bosons) can be
derived by adding one extra \suu\ invariant dimension-six
interaction term to the standard model Lagrangian.
This extention of the BKS model provides a mechanism which protects LEP I
observables against (strong) deviations from their standard model values.

In Sect.\ 2, we explain the KMSS model and the reduction to a single parameter
effective Lagrangian (BKS model).
In Sect.\ 3 we give an \suu\ invariant
derivation of the model and discuss its sensitivity to LEP I experiments.
In Sect.\ 4, the calculated cross sections
are presented and
discussed. We show the dependence of the cross sections on $\kp$.
A fit of the free parameter $\kappa$ on the basis of
the expected luminosity  and systematic error of future data to be taken
at the NLC is performed.
Sect.\ 5 contains our final conclusions.


\section[ ]{The effective Lagrangian}
As mentioned, our investigations are based on the KMSS model for
vector-boson self interactions \cite{schi1,schi2}. The effective Lagrangian is
constructed in the following way:
\begin{enumerate}
\item One starts with a \suwi\ triplet field $\vec{W}_\mu$
and an (unphysical) photon field $\tilde{A}_\mu$. Kinectic terms and a mass
term
for the \bfw\ field are introduced.
\item The most general
self interaction of \wmu\
in conformity with global \suwi\ symmetry and the restriction
to dimension-four couplings is constructed.
\item The photon field $\tilde{A}_\mu$ is coupled
to the \wmu\
field by minimal substitution in the \wmu\ kinetic term.
This assures electromagnetic gauge invariance and breaks global \suwi .
\item Mixing between the neutral weak vector boson $W_{3\mu}$
and $\tilde{A}_\mu$ is added.
\item Finally, an anomalous magnetic-moment term for the
interaction of the electromagnetic field with the charged vector bosons
is added.
\end{enumerate}
The Lagrangian obtained by this procedure is given in \cite{schi1}. It has four
free parameters (of which only two affect the trilinear couplings).

The next step is the analysis of the vector-boson scattering processes
$\rm V_1,V_2\to V_3,V_4$ with $V_i=W^\pm,Z$ performed in \cite{schi2}.
The requirement
of vanishing of the most strongly unitarity violating terms
of order $s^2$ in the tree amplitudes (which is eqivalent to the demand of
vanishing quartic divergences in one-loop corrections \cite{schi5})
leads to
three conditions on the four free parameters, so they can be expressed in
terms of a single one,
namely the $\rm W^\pm$ anomalous magnetic moment $\kappa$.
The (higgsless) standard model is obtained by choosing
$\kappa =1$.

The final Lagrangian ${\cal L}_{SI}$ for the self interactions is then given by
\begin{eqnarray}{\cal L}_{SI}&=&ie\left[A_\mu(W^{-\mu\nu}W_\nu^+-
W^{+\mu\nu}W_\nu^-)+\kappa F_{\mu\nu}W^{+\mu}W^{-\nu}\right]\nn
&+&ie\frac{\kappa-\sin^2\tw}{\sin\tw\cos\tw}
\left[Z_\mu(W^{-\mu\nu}W_\nu^+-
W^{+\mu\nu}W_\nu^-)+\frac{\kp\cos^2\tw}{\kp-\sin^2\tw}
Z_{\mu\nu}W^{+\mu}W^{-\nu}\right]\nn
&-&e^2(A_\mu A^\mu W_\nu^+W^{-\nu}-A_\mu A_\nu W^{+\mu}W^{-\nu})\nn
&-&2e^2\frac{\kappa-\sin^2\tw}{\sin\tw\cos\tw}\left[
A_\mu Z^\mu W_\nu^+W^{-\nu}-\frac{1}{2}A_\mu Z_\nu (W^{+\mu}W^{-\nu}
+W^{-\mu}W^{+\nu})\right]\nn
&-&e^2\frac{(\kappa-\sin^2\tw)^2}{\sin^2\tw\cos^2\tw}
(Z_\mu Z^\mu W_\nu^+W^{-\nu}-Z_\mu Z_\nu W^{+\mu}W^{-\nu})\nn
&+&\frac{1}{2}e^2\frac{1}{\sin^2\tw}\kp^2(W_\mu^-W_\nu^+W^{-\mu}W^{+\nu}-
W_\mu^-W_\nu^+W^{+\mu}W^{-\nu})
\label{bks}\end{eqnarray}
with $F_{\mu\nu}=\partial_\mu A_\nu-\partial_\nu A_\mu$,
$W^\pm_{\mu\nu}=\partial_\mu W^\pm_\nu-\partial_\nu W^\pm_\mu$, etc.,
and a single
free parameter $\kp$. Considering small deviations from the \suu\ value of
$\kp=1$, \be \kp\equiv 1+\Delta\kp,\qquad \Delta\kp\ll 1,\label{small}\ee
the last two terms in \eref{bks} may be approximated by terms linear in $\dk$,
i.\ e.,
\bea&-&e^2\frac{\cwq}{\sin^2\tw}\left(1+\frac{2\dk}{\cwq}\right)
(Z_\mu Z^\mu W_\nu^+W^{-\nu}-Z_\mu Z_\nu W^{+\mu}W^{-\nu})\nn
&+&\frac{1}{2}e^2\frac{1}{\sin^2\tw}(1+2\dk)(W_\mu^-W_\nu^+W^{-\mu}W^{+\nu}-
W_\mu^-W_\nu^+W^{+\mu}W^{-\nu})\label{linapprox}\eea
The Lagrangian \eref{bks} is a single-parameter extension of
the standard model Lagrangian and a reasonable reduction of the multi-parameter
Lagrangian of a model with arbitrary self interactions, since it is constructed
using the abovementioned  physical considerations, and it is the most general
Lagrangian in this formalism in which
the vector-boson scattering amplitudes do not contain $s^2$ terms.
It can be used to analyze
how changes of vector-boson self couplings {\em in terms of a single
free parameter\/}
influence the cross sections and to carry out a parameter fit.


\section[ ]{Derivation of the Model from an $\rm SU(2)_{L}
\times U(1)_{Y}$-invariant Lagrangian via the Higgs
mechanism\protect\footnote{We thank the referee of this paper whose questions
initiated the present section.}}
As mentioned,
the interactions \eref{bks} of the vector bosons with one another
are such that tree-level scattering amplitudes do not contain any
(tree-)unitarity-violating
terms growing as $s^2$ at high energies. Corresponding to
the absence of such terms, there are no $\Lambda^4$ divergences at the one loop
level
\cite{schi5}.
For the standard-model value of $\kp=1$, the remaining linear high-energy
growth of the scattering amplitudes (proportional to $s$)
is compensated by adding the Higgs
scalar with sufficiently low mass $M_H$. Vector-boson loops depend
logarithmically on $M_H$.

One may pose the question wether a simple convergence-producing mechanism also
exists in case of the Lagrangian \eref{bks} for $\kp\ne 1$. In this section we
will show, that the quadratic loop divergences can indeed be removed.
Addition of suitable non-standard
interactions of the Higgs scalar with the vector bosons
allows to embed the self interactions
\eref{bks} into an \suu\ symmetric framework. Even though the added
Higgs-scalar--vector-boson interactions involve nonrenormalizable dimension-six
interactions, they will nevertheless provide a sufficiently decent behaviour of
the loop corrections to protect LEP I observables from violent deviations from
standard predictions.

The \suu\ invariant Lagrangian, which yields the vector boson interactions
\eref{bks}, is
\be \lag_{SI}=\lag_{SM}+\dk\frac{g}{M_W^2}\owp\, ,\label{lsi}\ee
where $\dk=\kp-1$ as in \eref{small},
$\lag_{SM}$ denotes the standard-model Lagrangian, $g=e/\sw$
and $\owp$ is given by
\be {\cal O}_{W\Phi}=i(D_\mu\Phi)^\dagger\vec{\tau}\cdot\vec{W}^{\mu\nu}
(D_\nu\Phi)=-i\frac{v^2}{2}{\rm tr}(W_{\mu\nu}(D^\mu U D^\nu U^\dagger))
\label{owp}\ee
with $\vec{W}_{\mu\nu}=\partial_\mu\vec{W}_\nu-\partial_\nu\vec{W}
_\mu
-g\vec{W}_\mu\times \vec{W}_\nu$,
$W_{\mu\nu}=\frac{\vec{\tau}}{2}\cdot
\vec{W}_{\mu\nu}$, $U=\frac{\sqrt{2}}{v}
\Big(\tilde{\Phi}:\Phi\Big)$,$\tilde{\Phi}=i\tau_2\Phi^\ast$
and $D_\mu U = \partial_\mu U +ig\frac{\vec{\tau}}{2}
\cdot\vec{W}_\mu U-ig'U\frac{\tau_3}{2}B_\mu$.
Explicitly, upon introducing the $Z$  and photon fields via the usual
diagonalization, $\owp$ is given by

\newpage
\bea \frac{g}{M_W^2}\owp &=&
ie F_{\mu\nu}W^{+\mu}W^{-\nu}\nn
&+&ie\frac{1}{\sin\tw\cos\tw}
\left[Z_\mu(W^{-\mu\nu}W_\nu^+-
W^{+\mu\nu}W_\nu^-)+\cwq
Z_{\mu\nu}W^{+\mu}W^{-\nu}\right]\nn
&-&2e^2\frac{1}{\sin\tw\cos\tw}\left[
A_\mu Z^\mu W_\nu^+W^{-\nu}-\frac{1}{2}A_\mu Z_\nu (W^{+\mu}W^{-\nu}
+W^{-\mu}W^{+\nu})\right]\nn
&-&2e^2\frac{1}{\sin^2\tw}
(Z_\mu Z^\mu W_\nu^+W^{-\nu}-Z_\mu Z_\nu W^{+\mu}W^{-\nu})\nn
&+&e^2\frac{1}{\sin^2\tw}(W_\mu^-W_\nu^+W^{-\mu}W^{+\nu}-
W_\mu^-W_\nu^+W^{+\mu}W^{-\nu})\nn
&+&\mbox{Higgs couplings.}\label{owpev}\eea
Substituting this explicit form of $\owp$ into \eref{lsi}, one  easily verifies
that the gauge-boson self interactions contained in \eref{lsi} are identical to
those of the BKS Lagrangian \eref{bks} with the linear approximation
\eref{linapprox} for the quadrilinear interactions.
In the present paper we  only consider small values of $\dk$, so we can
identify the BKS self couplings \eref{bks} with those of
the \suu\ invariant model \eref{lsi}.

The effect of boson loops on LEP I observables due to
non-standard
interactions such as $\owp$ and other dimension-six terms was investigated in
\cite{deruj,hagzep}. Typically, in distinction from the \suu\ spontanously
broken standard model, for  $\owp$ one finds
a quadratic dependence on the Higgs-boson mass,
$M_H^2$, and a logarithmic dependence on the neessary cut-off,
$\ln\Lambda$.
Anomalous couplings of the kind of extra dimension-six terms
considered here are not
restricted very much by LEP I data, as these data are not very sensitive to
logarithmic cut-off-dependent loop corrections \cite{hagzep}.

In summary, the BKS Lagrangian, originally derived from global $\rm SU(2)_{WI}$
weak isospin symmetry broken by electromagnetism and vanishing of the most
strongly (as $s^2$) rising contribution to vector-boson-scattering amplitudes,
can be embedded into an \suu\ symmetric theory of the simple and compact form
\eref{lsi} with (dimension-six) non-standard
Higgs self interactions. This embedding provides an example of how non-standard
trilinear and quadrilinear vector-boson-scattering amplitudes can coexist with
``standard'' empirical LEP I results.

In the calculations of the present paper the Higgs sector of the theory is
disregarded (although it contributes to three-gauge-boson-production processes)
since we only want to study the effect of anomalous gauge-boson self
interactions. This is justified if we assume that $M_H$ lies above the energy
region to be investigated in \ep\ annihilation at the NLC\footnote{So our
cross sections for $\ws=2000\gev$ should not be taken too seriously, as they
would eventually be changed by the Higgs effects.} so that the contributions of
Higgs exchange to the cross sections are negligible.


\section[ ]{Cross-Sections}
Figure \ref{feyn} shows schematically the
tree-level Feynman diagrams which contribute
to the three-vector-boson production processes \wwz\ and \wwa\footnote{Except
for the Higgs boson contribution which we do not consider here as explained
above. For Higgs boson effects see \cite{bh,tnk}.}.
Altogether there are 15 diagrams. We
calculated the cross-sections in the same
way as Barger, Han and Phillips \cite{bh} evaluating the amplitudes
by employing
helicity techniques \cite{hz} and integrating numerically over the phase space.
All standard-model couplings were derived via the 1-loop relations
among the masses and coupling constants in the standard model
(e.g. see \cite{kks}) from the
Z-boson mass $M_Z$, the Fermi coupling $G_F$
and the electromagnetic fine structure constant
$\alpha_{\rm em}$. The latter  was set equal to $\alpha_{\rm em}=1/128.8$,
which is the high energy value
of the running coupling constant.
As in \cite{bh} we imposed the following transverse-momentum and
pseudorapidity cuts for the photon produced in the reaction \wwa :
\begin{equation} p_{t,\gamma}>20\gev,\quad |\eta_\gamma|<2\, .\end{equation}

First we study the total cross-sections for the production of unpolarized
vector bosons in the reactions \wwz\ and \wwa .  Figure \ref{tcs} shows these
cross-sections
as a function of the energy for different values of the free
parameter $\kp$ around the standard-model value of $\kp=1$.
We find with $\kp=1$ at $\ws=500\gev$ the results $\st=39\fb$ for \wwz\ and
$\st=135\fb$ for \wwa.
Beyond the threshold region standard model cross-sections
decrease with increasing
energy as a consequence of gauge cancellations, i.e., the
cancellations of those parts of the
amplitudes for the production of longitudinally polarized vector bosons
which grow as nonegative powers of
$s$. These
are caused by the relations among the self couplings
of the vector bosons and the couplings to fermions in the standard
model\footnote{In
distinction to vector-boson scattering, in these processes the Higgs boson
is not needed for good high energy behavior.}. If the free
parameter $\kp$ is set
$\kp\ne 1$ these relations are violated. Therefore, in those cases
there are no complete cancellations, and $\st$ is growing
with energy, so that  unitarity is violated
at high energies and must eventually be restored by ``new physics''
contributions.
That is why the deviations of the cross sections
in the general BKS model from the standard model increase with
the energy $\ws$ and with
$\dk$, which characterizes the magnitude of the deviations of the
self coupling constants from the standard model values.
Although we find  differences of some orders of magnitude
from the standard model
at the TeV energy scale, at the NLC energy of $\ws=500\gev$ there are just
some \% differences for our choices of $\kappa$ from 0.9 to 1.1. We will
come back to this point later.

To illustrate the effect of violation of the gauge cancellations, we show
the cross-sections for the production of exclusively transversely
and of exclusively longitudinally  polarized vector bosons in the reaction
\wwz\
(Fig. \ref{pol}).
Production of tranverse vector bosons
yields a large contribution to the total cross section
in the standard model.
However, for $\kp\ne 1$ the deviations from the \suu\ predicts
are very small, because
the amplitudes of the different Feynman graphs do not grow with energy and
no cancellations are necessary. In contrast, for the production of longitudinal
vector bosons, the standard model cross section is very small, but the
non-cancellation of the leading amplitudes
leads to enormous deviations, when $\kp$ departs from $\kp =1$.

We turn to the question of how well  trilinear and quadrilinear
self interactions will be measurable in future experiments. We determine the
empirical limits which can be assigned to $\kp$ if the standard model cross
sections will be confirmed in experiments. First,
we perform an analysis  at the NLC
energy of $\ws=500 \gev$. Our investigations are based on the
cross-sections \wwz , \wwa\ (production of unpolarized vector bosons)
and \wwzlll\ (production of longitudinal vector bosons).
The expected luminosity of NLC is $20\, \rm fb^{-1}a^{-1}$, so there would be
800 annual events of \wwz . Following the analysis of \cite{bh}, 20\% of them
will be reconstructable from the decay products of the final vector bosons.
If we determine the  statistical error to 90\% confidence level and assume
2\% systematic error we find a total error of 10\% after three years of
collecting data.
For the reaction \wwa\ the statistical error is smaller because of the
larger cross-section. By the same reasoning we find a total error of
5\% after three years of running. For the process \wwzlll\ the cross-section
of 0.5$\fb$ is extremely small and causes a large statistical error. The total
error in this case is 70\%.
As an outlook we perform the same analysis at the energy of
$\ws=2000\gev$
in order to see, how the precision
of the parameter fit increases with energy\footnote{We are aware of the fact
that these estimates should eventually be refined by taking into account the
effects of a Higgs scalar of unknown mass $M_H\le 2\,\rm TeV $.
Our results have to be considered as crude exploratory values.}.
Since no precise parameters of a 2000$\gev$ machine are available, we assume an
experimental error of 10\% for production of unpolarized vector bosons and
again 70\% error for the reaction \wwzlll .
 Figure \ref{par} shows the total cross
sections for these processes at the two abovementioned fixed energies as a
function of $\kp$.
We see, how small deviations of $\kappa$ from its standard model value of
$\kp=1$
affect the total cross-sections.
 From these results, we now can find
the interval around the standard model value of $\kp=1$ to which
$\kp$ is restricted if the deviations from the standard model  do not
exceed the experimental error.
The results of these parameter fits are given in table 1.
We see that NLC results can restrict $\kappa$ to a region of a few \%
around its standard model value of $\kp=1$. Accuracy increases by one
order of magnitude if $\ws$ is raised from 500$\gev$ to 2000$\gev$.
The production of exclusively longitudinal vector bosons
\wwzlll\ yields limits in the same order as production of unpolarized
gauge bosons, because the effect of the larger deviations from the
standard model is compensated by the larger experimental error.
\begin{table}[h]
\begin{center}
\begin{tabular}{|l|c|c|}
\hline Process&$\ws=500\gev$ (NLC)&$\ws=2000\gev$\\\hline\hline
\wwz & $0.95\le\kp\le 1.06$&$0.997\le\kp\le 1.006$\\ \hline
\wwa & $0.98\le\kp\le 1.09$&$0.997\le\kp\le 1.009$\\\hline
\wwzlll & $0.98\le\kp\le 1.07$&$0.999\le\kp\le 1.003$\\ \hline
\end{tabular} \end{center}
\caption[ ]{\label{tab}Results from the fit of the free parameter
$\kp$ based on the total cross-sections for \wwz , \wwa\ and \wwzlll\
at $\ws=500\gev$ and
$\ws=2000\gev$.}
\end{table}


\section[ ]{Conclusions}
\begin{itemize}
\item The three vector-boson production processes \wwz\ and \wwa\ supply
the phenomenological easiest way to test the tri- {\em and} quadrilinear
self interactions of the electroweak vector bosons directly.
\item The BKS model reduces the most general form of vector-boson
self interactions by some physically resonable assumptions to a single
parameter model, which can be used to study the sensitivity of the
standard model
cross sections to variations of the self couplings.
\item The BKS model is not restricted very much by present LEP I data since its
anomalous vetor-boson self couplings can be obtained from a
simple one-parameter locally \suu -invariant interaction
term. As a consequence, loop-contributions in this model
(extended by appropriate Higgs-couplings) diverge at most logarithmically.
\item The cross-sections of the three-gauge-boson-production processes
are very sensitive to variations of the free
parameter. At the NLC energy of $\ws=500\gev$ variations of $\kp$ at the
per-cent level lead to measurable differences from the standard model.
Measurement of the polarisation of the final vector bosons
yields only slighly stricter limits on $\kp$.
\item Although $\kp$ can be determined in \ww , even with stricter parameter
limits due to better statistics (see \cite{schi4} Table 4.2),
this does not imply the structure of the quadrilinear couplings.
Indeed, the agreement of the values of $\kappa$ deduced from \ww\ and
\wwz ,$\gamma$ measurements is essential for establishing the full
non-Abelian Yang--Mills structure.
\item In summary: Future \ep\ colliders at 500$\gev$ energy or more will supply
good empirical possibilities to explore the
self interactions of the vector bosons
and to test the Yang--Mills structure of the electroweak standard model.
\end{itemize}




\section*{Figure Captions}
\newcounter{fig}
\newlength{\fig}
\newlength{\numnum}
\settowidth{\numnum}{\bf 1}
\settowidth{\fig}{\bf Figure 1:}
\begin{list}{\bf Figure \makebox[\numnum][r]{\arabic{fig}}:}{\usecounter{fig}
\labelwidth\fig \leftmargin\labelwidth \addtolength\leftmargin\labelsep}
\item\label{feyn}Feynman diagrams for three-vector-boson production
\item\label{tcs}Total cross-sections for the reactions (a) \wwz\
and (b) \wwa\ as a function of the CM-energy for different
values of $\kp$. The solid lines show the results for the standard-model
value of $\kp=1$.
\item\label{pol}Cross-sections for the production of (a)
exclusively transversely
polarized vector bosons $\rm e^+e^-\to W^+_TW^-_TZ^{ }_T$ and  (b) exclusively
longitudinally polarized vector bosons $\rm e^+e^-\to W^+_LW^-_LZ^{ }_L$ as a
function of $\ws$ for different $\kappa$.
\item\label{par}Total cross-sections for (a) \wwz ,  (b) \wwa\ and (c) \wwzlll\
for fixed
$\ws=500\gev$ and $\ws=2000\gev$ as a function of $\kp$
(solid lines). The dashed lines border the regions where the results
agree with the standard model value within the estimated experimental error.
\end{list}

\begin{thebibliography}{99}
\frenchspacing
\bibitem{sm}S. L. Glashow, Nucl. Phys. B22 (1961) 579;\\
S. Weinberg, Phys. Rev. Lett. 19 (1967) 1264;\\
A. Salam, Proc. 8th Nobel Symposium, ed. N. Svartholm (Almquits and Wiksells,
Stockholm, 1968) p. 367
\bibitem{nlc}``\ep\ Collisions at 500$\gev$, the
Physics Potential'', ed. P. W. Zerwas, DESY 92-123
\bibitem{gg}K. Gaemers and G. Gounaris, Z. Phys. C1 (1979) 259;\\
K. Hagiwara, R. D. Peccei and D. Zeppenfeld, Nucl. Phys. B282 (1987) 253
\bibitem{schi4} G. Gounaris, J. L. Kneur, J. Layssac, G. Moultaka, F. M. Renard
and D. Schildknecht, in \cite{nlc} p.\ 735
\bibitem{schi1}M. Kuroda, J. Maalampi, K. H. Schwarzer and D. Schildknecht,
Nucl. Phys. B284 (1987) 271
\bibitem{schi2}C. Bilchak, M. Kuroda and D. Schildknecht, Nucl. Phys.
B299 (1988) 7
\bibitem{schi3}M. Kuroda, F. M. Renard and D. Schildknecht, Z. Phys. C40 (1988)
575
\bibitem{bh}V. Barger, T. Han and R. J. N. Phillips, Phys. Rev. D39 (1989) 146
\bibitem{tnk}A. Tofighi-Niaki and J. F. Gunion, Phys. Rev. D39 (1989) 720
\bibitem{wwz}
F. M. Renard, Z. Phys. C2 (1979) 17;\\
E. N. Agryresi, O. Karakianitis, C. G. Papadopoulos and W. J. Stirling
Phys. Lett. B259 (1991) 195;\\
W. Benakker, F. A. Berends and  T.Sack,
Nucl. Phys. B367 (1991) 287;\\
J. Kalinowski, in \cite{nlc} p.\ 217;\\
G. B\'{e}langer and F. Boudjema, in \cite{nlc} p.\ 783;\\
G. Cveti\v{c}, C. Grosse-Knetter and R. K\"ogerler, in \cite{nlc} p.\ 775 and
Bielefeld-Preprint BI-TP 92/32, submitted to Nucl. Phys B
\bibitem{schi5}H. Neufeld, J. D. Stroughair and D. Schildknecht,
Phys. Lett.  B198 (1987) 563
\bibitem{deruj} A. de R\'{u}jula, M. B. Gavela, P. Hernandez and E. Mass\'{o},
Nucl. Phys. B384 (1992) 3
\bibitem{hagzep}K. Hagiwara, S. Ishihara, R. Szalapski and D. Zeppenfeld,
Phys. Lett. B283 (1992) 353
\bibitem{hz}K. Hagiwara and D. Zeppenfeld, Nucl. Phys. B274 (1986) 1
\bibitem{kks}J. L. Kneur, M. Kuroda and D. Schildknecht, Phys. Lett. B262
(1991) 93
\end{thebibliography}
\end{document}